\documentclass[12pt]{article}
\usepackage{paper2e}
\usepackage{mydefs2e}
\renewcommand{\Box}{\,\raisebox{-.45pt}{\drawsquare{7}{0.6}}\,}
\newcommand{\MP}{M_{\rm P}}
\renewcommand{\d}{\partial}
\begin{document}

\begin{titlepage}\preprint{CERN-TH/2003-044\\
UMD-PP-03-041}

\title{Strong Interactions and Stability\\\medskip
in the DGP Model}

\author{Markus A. Luty}
\address{Department of Physics, University of Maryland\\
College Park, 
MD 20742, USA}

\author{Massimo Porrati}

\address{Department of Physics, New York University\\
4 Washington Place, New York, NY 10012, USA}

\author{Riccardo Rattazzi\footnote{On leave from INFN, Pisa, Italy.}}

\address{Theory Division, CERN\\
CH-1211, Geneva 23, Switzerland}

\begin{abstract}
The model of Dvali, Gabadadze, and Porrati (DGP) gives a simple
geometrical setup in which gravity becomes 5-dimensional
at distances larger than a length scale $\la_{\rm DGP}$.
We show that this theory has strong interactions at a length
scale $\la_3 \sim (\la_{\rm DGP}^2 / \MP)^{1/3}$.
If $\la_{\rm DGP}$ is of order the Hubble length, then
the theory loses predictivity at distances shorter than
$\la_3 \sim 1000$~km.
The strong interaction can be viewed as arising from a 
longitudinal `eaten Goldstone' mode that gets a small kinetic term only
from mixing with transverse graviton polarizations, analogous to the case
of massive gravity.
We also present a negative-energy classical solution, which can be
avoided by cutting off the theory at the same scale
scale $\la_3$. 
Finally, we examine the dynamics of the longitudinal Goldstone mode when
the background geometry is curved. 
\end{abstract}

\end{titlepage}

\section{Introduction}
The DGP model \cite{DGP} is the first ghost-free example
of a mechanism in which gravity can be localized on a 4D brane
in a space of infinite transverse volume.
It describes a theory where 4D general
covariance is unbroken, yet the graviton is a metastable state. Its main 
property is that, on the 4D brane, gravity looks 4D at short distance, 
while it weakens at large distance. This property suggest an interesting 
alternative to the standard description of our present-day accelerating 
universe. In DGP, the cosmic acceleration could be 
due to gravity becoming weaker
at large (horizon-size) distance, rather than to a positive cosmological 
constant. Explicit realizations of this scenario have been proposed, for 
instance in \Ref{DDG}.   

The model can be described by the action
\beq\bal
S_{\rm DGP} = 2 & M_5^3 \int_{\scr{M}} d^5 x\, \sqrt{-G}\, R(G)
\\
&+ \int_{\d\scr{M}} d^4 x\, \sqrt{-\ga} \left[
-4 M_5^3 K(\ga) + 2 M_4^2 R(\ga) \right],
\eal\eeq
where $\scr{M}$ is a 5D manifold with boundary $\d\scr{M}$,
$G$ is the 5D metric, $\ga$ is the 4D induced metric on the boundary,
and $K$ is the extrinsic curvature.%
\footnote{The boundary can also be treated as an orbifold fixed point.
We will discuss the relation between these approaches below.}
In this model gravity on the brane looks 4D at distances
shorter than
\beq
\la_{\rm DGP} = 
\frac{M_4^2}{M_5^3}.
\eeq
For $M_4 \gg M_5$ this can be a macroscopic length,
for example the size of the present horizon.

The DGP model is closely related to massive gravity.
In fact, the brane-to-brane graviton propagator can be written
\beq
D_{\mu\nu\,\rho\si}^{\rm DGP}(p)
= D_{\mu\nu\,\rho\si}^{\rm massive}(p, |p|/\la_{\rm DGP}),
\eeq
where $D^{\rm massive}_{\mu\nu,\la\rho}(p, m^2)$ is the propagator for
4D massive gravity.
The DGP therefore shares with massive gravity the `vDVZ discontinuity'
\cite{vdvz}:
at distances smaller than $\la_{\rm DGP}$, the model reduces not to
general relativity, but to a
scalar--tensor theory of gravity, where the scalar couples with
gravitational strength (it does not decouple
in the limit $m \to 0$).
\Refs{vdvz} showed that in the one graviton exchange approximation,
massive gravity predicts unacceptable deviations in the predicted
bending of light by the sun.

However, as shown by Vainshtein \cite{v} for massive gravity,
the situation is actually more subtle.
Near a heavy source, the
one graviton exchange approximation breaks down at very large
distances, and he argued that at smaller distances the resummation
of nonlinear effects \emph{restores} agreement with general
relativity.
In DGP, the one graviton exchange approximation breaks down at
distances $R_* \sim (R_{\rm S} \la_{\rm DGP}^2)^{1/3}$, where
$R_{\rm S}$ is the Schwarzschild radius of the source.
At smaller distances, it was explicitly shown that that the full
nonlinear solution approaches that of general relativity \cite{ddgv,g,p3}.
Because the scale $R_*$ is very large for astrophysical sources,
it appeas that the DGP model may describe
our universe \cite{DDG}. 

The fact that the one-particle 
exchange breaks down at a distance so much larger than $R_{\rm S}$ suggests
that DGP has hidden strong interaction scales.
For a large classical source with $R_{\rm S} \gg 1/M_4$
the non-linearities at the scale $R_*$ can certainly be associated to
classical physics.
On the other hand, for a source with $R_{\rm S}\sim 1/ M_4$, corresponding
heuristically to one quantum of gravitational charge,
we expect any non-linearity to be due to quantum physics.
Based on this qualitative argument we expect
strong quantum effects to become important at a length scale
\beq[DGPstrong]
\la_3 = 
\left( \frac{\la_{\rm DGP}^2}{M_4} \right)^{1/3}.
\eeq
In this paper we show that this is precisely what happens.
For $\la_{\rm DGP}$ of order the Horizon size, $\la_3 \sim 1000$~km.
At distances shorter than $\la_3$, new interactions become important,
and there seems to be no reason that the theory should agree with
general relativity.

An analogous strong interaction is also present
in massive gravity.
The strong interactions can be made manifest
using the St\"uckelberg trick of
nonlinearly realizing the gauge invariance broken by the
mass term \cite{AGS}.
In massive gravity, the St\"uckelberg (or Goldstone) fields have strong
self-interactions that necessitate a cutoff that goes to zero as
the graviton mass goes to zero.
An analogous phenomenon is familiar for massive non-Abelian gauge fields,
where the St\"u{}ckelberg sector is 
a non-linear sigma model that is strongly interacting at a scale $m / g$,
where $m$ is the gauge boson mass and $g$ is the gauge coupling.
In the case of massive gravity, \Ref{AGS} showed 
that the theory becomes strongly interacting at a scale
$\La_5 \sim (m^4 \MP)^{1/5}$, or $\La_3 \sim (m^2 \MP)^{1/3}$
if the leading strong terms are tuned to be small.
Substituting the `running mass' $|p| / \la_{\rm DGP}$ into the
$\La_5$ cutoff for massive gravity and solving for $p$
also suggests that the DGP model
has strong interactions at the scale \Eq{DGPstrong}.

In this paper, we study the DGP model in detail to rigorously
establish the existence of the strong interactions and understand
their origin.
Following the logic of the St\"uckelberg trick,
we introduce extra pure gauge degrees of freedom to
parameterize the strong interactions.
We do this by formulating the theory on a space with boundary, with no
{\it a priori} boundary conditions on the fields.
This reduces to conventional orbifold boundary conditions
in a particular gauge, but a different gauge choice is useful to
make the strong interactions manifest.
Since DGP is a generally covariant theory, it is not surprising that
we find that the St\"u{}ckelberg mode has a geometrical
interpretation: it is a `brane-bending' mode that keeps the
induced boundary metric fixed.

We also find evidence for strong interactions at the scale
\Eq{DGPstrong} at the classical level.
We show that the DGP model has classical solutions with negative
5D energy, with a boundary stress tensor obeying the dominant
energy condition.
These solutions are at the edge of the regime of validity of the
effective theory with short-distance cutoff given by \Eq{DGPstrong},
giving another indication of new physics at that
scale.

We then consider the behavior of the DGP model in the presence of
curvature.
We show that for the case of a positive curvature boundary (de Sitter sign),
the self-interactions become stronger, and the  Goldstone mode
becomes a ghost for sufficiently large curvature.
Closely related results have been found for massive gravity in
\Refs{p}.
We also consider the Randall-Sundrum model \cite{rs}
with a DGP kinetic term.
For a DGP kinetic term on the Planck brane, we find no strong interactions,
in agreement with expectations based on holography.
For a DGP kinetic term on the IR brane, the radion becomes a ghost if
$\la_{\rm DGP}$ becomes larger than the 5D AdS length.

This paper is organized as follows.
Section 2 explains the boundary effective action we will use as a tool
for the case of a toy scalar model.
Section 3 uses this formalism to compute the boundary action for
the DGP model.
We find an effective action for the St\"u{}ckelberg mode and explicitly
compute the cubic interactions.
We discuss the power counting and derive a non-renormalization theorem
for the cubic interaction.
Section 4 describes a negative energy solution.
Section 5 extends the analysis of section 3 to the case of backgrounds
with nonzero curvature.

\section{Boundary Effective Action}
In this section, we describe the formalism we use to obtain an
effective action for the boundary field in a theory such as DGP.
Consider a field theory on a space with boundary, and
suppose that we are interested in the correlation functions
of sources on the boundary.
We do not impose any {\it a priori} boundary conditions
on the fields.
In the path integral, we integrate over arbitrary boundary values of the
bulk fields weighted by their action.

It is useful to separate the fields into bulk fields $\Phi$ and boundary
fields $\phi$.
Locality of the action means that the path integral can be written as
\beq
Z = \myint d[\Phi]\, d[\phi]\, e^{i \left( S_{\rm bulk}[\Phi]
+ S_{\rm bdy}[\phi] \right)}.
\eeq
Since the only sources are on the boundary 
we can integrate out the bulk fields to obtain a (nonlocal)
effective action for the boundary fields.
We must therefore integrate over all $\Phi$ with boundary condition
\beq
\Phi| = \phi,
\eeq
where `$|$' indicates evaluation at the boundary.
We perform the $\Phi$ integral semi-classically,
by expanding about a solution $\bar\Phi$ to the bulk equations of motion,
with $\left. \de \Phi \right| = 0$ because of the boundary condition.%
\footnote{Because $\left. \de \Phi \right| = 0$ there is no boundary
term in the variation of the bulk action.}
The semi-classical expression for the path integral is then
\beq
Z = \myint d[\phi]\, e^{i \left( S_{\rm bdy}[\phi]
+ \Ga[\phi] \right)},
\eeq
where the effective action from integrating out the bulk is
\beq
e^{i \Ga[\phi]} = 
e^{i S_{\rm bulk}[\bar\Phi]}
\myint d[\Phi']\,
\exp\left\{ \frac{i}{2} \myint \Phi'
\left. \frac{\de^2 S_{\rm bulk}}{\de \Phi^2} \right|_{\Phi = \bar{\Phi}}
\Phi' + \cdots \right\},
\eeq
where the path integral over $\Phi' = \Phi - \bar\Phi$ is performed over fields
with boundary condition
$\left. \Phi' \right| = 0$.

\subsection{Scalar Field Theory}
Let us do a simple example: free massless scalar field theory
in a 5D space with 4D boundary at $y = 0$.
The action is%
\footnote{If we had written the bulk kinetic term as
$\sfrac 12 \Phi \Box_5 \Phi$ there would be a boundary term in the
variation proportional to $\Phi \d_y (\de\Phi)$, affecting the behavior
of solutions near the boundary.}
\beq
S = \myint d^5 x \left[
- \sfrac 12 \d^M \Phi \d_M \Phi \right]
+ \myint d^4 x \left[ -\sfrac 12 \ka \d^\mu \phi \d_\mu \phi \right]_{y = 0},
\eeq
where
\beq
\left. \Phi \right| = \phi.
\eeq
The classical solution for $\Phi$ with these boundary condition is
\beq
\bar\Phi(x, y) = e^{-y \De} \phi(x),
\eeq
where $\De = \sqrt{-\Box_4}$.
We therefore obtain
\beq
\Ga[\phi] &= \myint d^5 x \left[ 
\sfrac 12 \bar\Phi \Box_5 \bar\Phi \right]
+ \myint d^4 x \left[ \sfrac 12 \bar\Phi \d_y \bar\Phi \right]_{y = 0}
\\
&= -\myint d^4 x \left[ \sfrac 12 \phi \De \phi \right].
\eeq
From this we can read off the propagator for the $\phi$ field:
\beq
\avg{\phi \phi} = \frac{1}{\ka \Box_4 - \De}.
\eeq

Now we add bulk interactions:
\beq
S_{\rm bulk} = \myint d^5 x \left[
- \sfrac 12 \d^M \Phi \d_M \phi - \sfrac 13 \la \Phi^3 \right].
\eeq
The classical bulk field satisfies
\beq
\Box_5 \bar\Phi - \la {\bar\Phi}^2 = 0,
\qquad
\bar\Phi | = \phi.
\eeq
We find the solution order by order in $\la$:
\beq
\bar\Phi = {\bar\Phi}_0 + \bar\Phi_1 + \cdots,
\eeq
where $\bar\Phi_n = \scr{O}(\la^n)$.
${\bar\Phi}_0$ was computed above.
Because $ {\bar\Phi}_0 | = \phi$, we have
\beq
\bar\Phi_n | = 0 \quad \hbox{\rm for\ } n \ge 1.
\eeq
We now compute the first-order correction to the action:
\beq
\Ga[\phi] = \Ga_0 + \Ga_1 + \cdots,
\eeq
where $\Ga_0$ was computed above, and
\beq\nonumber
\Ga_1 &= \myint d^5 x \left[ -\d^M \bar\Phi_1 \d_M \bar\Phi_0 
- \sfrac 13 \la \bar{\Phi}_0^3 \right]
\\
\nonumber
&= \myint d^5 x \left[ \bar\Phi_1 \Box_5 \bar\Phi_0
- \sfrac 13 \la \bar{\Phi}_0^3\right]
+ \myint d^4 x 
\left[ - \bar\Phi_1 \d_y \bar\Phi_0 \right]_{y = 0}
\\
\nonumber
&= -\sfrac 13 \la \myint d^5 x\, \bar{\Phi}_0^3
\\
&= -\sfrac 13 \la \myint \frac{d^4 p_1}{(2\pi)^4}
\cdots \frac{d^4 p_3}{(2\pi)^4}\,
(2\pi)^4 \de^4(p_1 + p_2 + p_3)
\, \frac{\tilde{\phi}(p_1) \tilde{\phi}(p_2) \tilde{\phi}(p_3)}
{\sqrt{p_1^2} + \sqrt{p_2^2}+ \sqrt{p_3^2}},
\eeq
where
\beq
\tilde{\phi}(p) = \myint d^4 x\, e^{ip\cdot x} \phi(x).
\eeq
\section{Boundary Effective Action for Gravity}
We now consider 5D gravity with a 4D boundary, the case of interest.
We use Latin capitals $M, N, \ldots = 0, \ldots, 3; 5$ for 5D
spacetime indices, and $\mu, \nu, \ldots$ for 4D ones.
We denote the bulk metric by $G_{MN}$.

In the standard treatment, we define boundary conditions by imposing
an orbifold projection under reflections about the boundary.
Here instead we will not impose any boundary conditions on $G_{MN}$.
This means that there is extra gauge freedom in this formulation.
In the bulk, we have infinitesimal gauge transformations generated
by $\Xi_M$:
\beq
\de G_{MN} = \Xi^P \d_P G_{MN} + \d_M \Xi^P G_{PN} + \d_N \Xi^P G_{MP}.
\eeq
In the orbifold formulation, we have $G_{5\mu}| = 0$ and hence
$\dot{\Xi}_\mu | = 0$, $\Xi_5| = 0$, where the dot denotes the
derivative with respect to $x^5$ and the vertical stroke
denotes evaluation
at the boundary.
In the present formulation, $G_{5\mu}| \ne 0$ and
$\dot{\Xi}_\mu | \ne 0$.
We still have $\Xi^5 | = 0$ because we  use coordinates where the
boundary position is fixed.
We can choose a gauge where $G_{5\mu}| = 0$ to recover the orbifold boundary
conditions, so this formulation is completely equivalent to the usual one.
However, the extra gauge degrees of freedom in the present approach
are very useful in uncovering the strong interactions, as for massive
gravity~\cite{AGS}.

It is convenient to make a $4 + 1$ split and write the action in terms
of ADM-like variables~\cite{ADM}:
the lapse $N = (G^{55})^{-1/2}$,
the shift $N_\mu = G_{5\mu}$,
and the 4D metric $\ga_{\mu\nu} = G_{\mu\nu}$
on surfaces of constant $y = x^5$:
\beq[LbulkADM]
S_{\rm bulk} = 2 M_5^3 \myint d^4 x \int_0^\infty dy\,
\sqrt{-\ga} N\left[
R(\ga) - K^{\mu\nu} K_{\mu\nu} + K^2 \right],
\eeq
where
\beq[Kdef]
K_{\mu\nu} = \frac{1}{2N}(\dot{\ga}_{\mu\nu} - D_\mu N_\nu - D_\nu N_\mu)
\eeq
is the extrinsic curvature.
Here, 4D indices are raised and lowered with $\ga_{\mu\nu}$,
$D_\mu$ is the covariant derivative with respect to the 4D
metric $\ga_{\mu\nu}$, and the dot denotes a derivative with respect
to $y$.
Note that only first derivatives appear in the action and
there is no boundary (Gibbons--Hawking) term in this formulation
\cite{ADM,GH} (see also~\cite{Wald}).

In order to integrate out the bulk fields we must choose a gauge
for them.
We want to choose a gauge such that the propagator has manifestly
good high-energy behavior, so we choose de Donder gauge.
We write
\beq
G_{MN} = \eta_{MN} + H_{MN},
\qquad
\ga_{\mu\nu} = \eta_{\mu\nu} + \zeta_{\mu\nu},
\eeq
and add the gauge fixing term
\beq[gf]
\scr{L}_{\rm bulk, gf} &= - M_5^3 F^M F_M,
\eeq
where
\beq
F_M = \d^N H_{MN} - \sfrac 12 \d_M H.
\eeq
Classically, this imposes the gauge $F_M = 0$.
In terms of the $4 + 1$ split,
\beq
\scr{L}_{\rm bulk, gf} = -M_5^3 
[
(\d^\mu \zeta_{\mu\nu} - \sfrac 12 \d_\nu \zeta
- \sfrac 12 \d_\nu H_{55} + \dot{N}_\nu)^2
+ (\d^\mu N_\mu + \sfrac 12 \dot{H}_{55} - \sfrac 12\dot\zeta)]^2
\eeq
This leaves residual gauge freedom under infinitesimal transformations
satisfying
\beq[res]
\Box_5 \Xi_M = 0.
\eeq
Because we require $\Xi_M$ to be well-behaved at infinity,
and $\Xi_5 | = 0$, we see that $\Xi_5$ is completely fixed but
there is a residual gauge freedom, parameterized by
$\xi_\mu = \Xi_\mu |$.
Explicitly, the residual gauge freedom is
\beq
\Xi_\mu = e^{-y \De} \xi_\mu.
\eeq
This residual gauge freedom acts on the boundary fields at
linear order as
\beq[bdygauge]
\de h_{\mu\nu} = \d_\mu \xi_\nu + \d_\nu \xi_\mu,
\quad
\de N_\mu = - \De \xi_\mu,
\quad
\de h_{55} = 0.
\eeq

We now integrate out the bulk fields to obtain the quadratic
boundary action.
We solve the bulk equations of motion with boundary conditions
\beq
\bar{H}_{MN} = h_{MN}.
\eeq
The equations of motion in de Donder gauge are
\beq
\Box_5 (\bar{H}_{MN} - \sfrac 12 \eta_{MN} \bar{H}) = 0,
\eeq
with solution
\beq
\bar{H}_{MN} = e^{-y \De} h_{MN}.
\eeq
The induced boundary action is
\beq[bound1]
\Ga &= M_5^3 \myint d^4 x \bigl[
- \sfrac 12 h^{\mu\nu} \De h_{\mu\nu}
+ \sfrac 14 h_4 \De h_4
+ \sfrac 12 h_4 \De h_{55} -\sfrac 14 h_{55}\De h_{55} 
\nonumber\\
& \qquad\qquad\qquad\quad
- N^\mu \De N_\mu
- N^\mu (\d_\mu h_4 + \d_\mu h_{55} - 2 \d^\nu h_{\mu\nu})
\bigr].
\eeq
This is invariant under the gauge transformations \Eq{bdygauge}.
In fact, in terms of the invariant combination
\beq
\tilde{h}_{\mu\nu} = h_{\mu\nu} + \frac 1\De (\d_\mu N_\nu + \d_\nu N_\mu)
= -\frac{1}{\De} K_{\mu\nu}
\eeq
we have
\beq
\Ga = M_5^3 \myint d^4 x \bigl[
-\sfrac 12 \tilde{h}^{\mu\nu} \De \tilde{h}_{\mu\nu}
+ \sfrac 14 \tilde{h}_4 \De \tilde{h}_4
+\sfrac 12 \tilde{h}_4 \De h_{55}
- \sfrac 14 h_{55} \De h_{55} \bigr].
\eeq

The induced boundary action must be added to the DGP kinetic term
on the boundary:
\beq
\scr{L}_{\rm bdy,DGP} = M_4^2 
\left[
-\sfrac 12 (\d_\mu h_{\nu\rho})^2
+ (\d^\mu h_{\mu\nu})^2
- \d^\mu h_4 \d^\nu h_{\mu\nu}
+ \sfrac 12 (\d_\mu h_4)^2 \right].
\eeq
We fix the remaining gauge freedom parameterized by $\xi_\mu$
by adding a gauge fixing term
\beq
\scr{L}_{\rm bdy, gf} = -M_4^2  
( \d^\mu h_{\mu\nu} - \sfrac 12 \d_\nu h_4 + m N_\nu)^2,
\eeq
where $m = M_5^3 / M_4^2$ is the DGP scale.
This gauge fixing makes the large DGP kinetic term invertible,
and also eliminates the mixing between $h_{\mu\nu}$ and $N_\mu$.
The complete quadratic boundary Lagrangian
is then
\begin{eqnarray}
\scr{L}_{\rm bdy} &=&
M_4^2 \bigl[ 
\sfrac 12 h^{\mu\nu} (\Box_4 - m \De) h_{\mu\nu}
- \sfrac 12 h_4 (\Box_4 - m \De) h_4
\nonumber\\
&&
- m N^\mu (\De + m) N_\mu
+ \sfrac 12 m h_4 \De h_{55}
-m N^\mu \d_\mu h_{55}
- m\sfrac 14 h_{55} \De h_{55}
\bigr]. 
\end{eqnarray}
To see the strongly interacting mode, we consider the scalar modes
\beq
h_{\mu\nu} = \phi \eta_{\mu\nu},
\quad
N_\mu = \frac{1}{\De} \d_\mu \si,
\eeq
and $h_{55}$.
In the regime $p \gg m$ the leading terms are
\beq
\scr{L}_{\rm bdy} \simeq M_4^2 \left[
-2 \phi \Box \phi + 2 m \phi \De h_{55}
- m (\si + \sfrac 12 h_{55}) \De (\si + \sfrac 12 h_{55}) \right].
\eeq
From this we see that there is one scalar mode that gets a kinetic
term only through mixing with $h_{\mu\nu}$.
This mode can be parameterized by
\beq[strongbdymode]
N_\mu = \d_\mu \pi,
\qquad
h_{55} = -2 \De \pi.
\eeq
We can diagonalize the full kinetic term by defining
\beq[diag]
N'_\mu = N_\mu - \d_\mu \pi,
\qquad
h'_{\mu\nu} = h_{\mu\nu} + m \pi \eta_{\mu\nu},
\eeq
and we obtain
\beq
\scr{L}_{\rm bdy} \simeq M_4^2 \left[
\sfrac 12 h'_{\mu\nu} \Box_4 h'_{\mu\nu}
- \sfrac 14 h'_4 \Box_4 h'_4
- m N'^\mu \De N'_\mu
+ 3 m^2 \pi \Box_4 \pi \right].
\eeq
The small coefficient of the $\pi$ kinetic term is the origin of
the strong interactions in this theory.

We can characterize the strongly-interacting mode in another way,
which makes the generalization to curved backgrounds more transparent.
The mode we found can be characterized by the following three
properties:
$(i)$ it solves the linearized bulk equations of motion;
$(ii)$ $H_{\mu\nu} = 0$;
$(iii)$ it obeys the de Donder gauge-fixing condition
\beq
F_N = \d^M H_{MN} - \sfrac 12 \d_N H = 0.
\eeq
To see this, note that $H_{5\mu}$ and $H_{55}$ can be
locally gauged away, so any configuration satisfying these
conditions must be pure gauge in the bulk at linear order.
In fact, the mode \Eq{strongbdymode} extended into the bulk is
\beq[mm2]
H_{\mu\nu} = 0,
\qquad
H_{5\mu} = \d_\mu \Xi_5,
\qquad
H_{55} = 2 \dot{\Xi}_5,
\eeq
where
\beq
\Xi_5 = e^{-y \De} \pi.
\eeq
Since $\Xi_5 | \ne 0$ this gauge transformation is not a symmetry
of the full action with boundary.
The boundary shifts under this transformation, so this can be viewed
as a `brane bending mode.'
This is the only nontrivial configuration with the three properties
described above.
Beyond linear order, $\Xi_5|$ and $\pi$ have different interactions,
since $\Xi_5|$ (unlike $\pi$)
affects the 4D induced metric at order $(\Xi_5|)^2$

In fact, one could have anticipated by purely geometrical and physical 
arguments which mode, if any, could interact strongly.
The strong mode should be related with the UV properties
at the boundary, so we expect it to correspond to a trivial bulk geometry away
from the brane.
This is to say that the mode should be pure gauge in the bulk.
Moreover, it should also not correspond to sizeable curvature of the 
induced boundary geometry: this is because the large DGP Einstein 
term disfavors intrinsic curvature.
Since both the bulk and brane geometry should not be excited, the only
remaining geometrical object that can be excited by the mode is the extrinsic
curvature of the boundary, describing its shape as seen by a 5D observer. 
Up to trivial 4D reparametrizations,
the only mode satisfying the above three requirements is precisely \Eq{mm2}.
Notice that the second requirement bears similarity to the case of 
massive non-abelian gauge theory.
There, the strongly interacting Goldstones are the the pure gauge 
configurations $A_\mu=U^\dagger\partial_\mu U$
for which the gauge kinetic term vanishes. 
The gauge kinetic term, whose coefficient $1/g^2$ in principle can be 
very large, is the analog of the DGP term.

We now turn to the question of higher-dimension operators in the effective
theory.
Bulk interactions with higher powers of $H_{MN}$
will give rise to boundary interactions
of the form
\beq
\De \scr{L}_{\rm bdy} \sim M_5^3 \d (N_\mu)^p (\d \pi)^q
\sim m M_4^2 \d \left( \frac{\hat{N}_\mu}{m^{1/2} M_4} \right)^p
\left( \frac{\d \hat\pi}{m M_4} \right)^q,
\eeq
where
\beq
\hat \pi \sim m M_4 \pi,
\qquad
\hat N_\mu \sim m^{1/2} M_4 N_\mu
\eeq
are the fields with unit kinetic term.
From this we can read off a strong interaction scale
\beq
\La_{(p, q)} \sim \left( m^{p/2 + q - 1}
M_4^{p + q - 2} \right)^{1/(3p/2 + 2q + 1 )}.
\eeq
The lowest scale occurs for $p = 0$, $q = 3$ (cubic $\pi$ interactions),
which gives a scale
\beq[strongscale]
\La \sim (m^2 M_4)^{1/3}.
\eeq
Higher derivative terms in the bulk give rise to terms with additional
powers of $\d / M_5$.
Since $M_5 \gg \La$, these will give weaker interactions.

Notice that cubic terms cannot be canceled by changing the gauge condition.
Suppose that we modify the gauge-fixing condition by adding non-linear
terms in $H_{MN}$: $F'_M = F_M + \scr{O}(H^2)$.
Since the mode \Eq{strongbdymode} 
obeys
$F_M=0$, the new gauge condition becomes $F'_M=O(\pi^2)$.
This change produces only terms of order $\pi^4$ or higher in the
boundary action.

We now show that the cubic terms are present by
computing them explicitly. 
We must evaluate the cubic terms in $\pi$ in the configuration
(see \Eq{strongbdymode})
\beq[Pisoln]
N_\mu = \d_\mu \Pi,
\qquad
H_{55} = 2 \d_5 \Pi,
\eeq
where
\beq
\Pi = e^{-y \De} \pi.
\eeq
To find the cubic terms in the bulk action \Eq{LbulkADM}
we need $N$ to linear order and $K_{\mu\nu}$ to quadratic order:
\beq
N &= 1 + \sfrac 12 H_{55}
\\
K_{\mu\nu} &= \sfrac 12 
(1 - \sfrac 12 H_{55}) (\d_\mu N_\nu + \d_\nu N_\mu).
\eeq
The cubic terms involving $\pi$ and $N_\mu$ are
\beq
\De\scr{L}_{\rm bulk} = M_5^3 \left[
\sfrac 14 H_{55} (\d_\mu N_\nu - \d_\nu N_\mu)^2
+ \d^\mu H_{55} N_\mu \d^\nu N_\nu
- \d^\nu H_{55} N_\mu \d^\mu N_\nu \right].
\eeq
Using \Eq{Pisoln} and $\De \Pi = -\dot{\Pi}$,
this can be written
\beq
\De\scr{L}_{\rm bulk} &= 2 M_5^3 (\d_\mu \De \Pi)
(\d^\mu \Pi \Box_4 \Pi - \d_\nu \Pi \d^\mu \d^\nu \Pi)
\nonumber\\
&= -M_5^3 \d_y \left[ \d_\mu \Pi \d^\mu \Pi \Box_4 \Pi \right].
\eeq
Integrating this solution over $y$, we obtain
\beq[3int]
\De \scr{L}_{\rm bdy} = M_4^2 m \d^\mu \pi \d_\mu \pi \Box_4 \pi.
\eeq
To see that this cubic interaction represents a physical effect,
we can compute the correlation function of three stress-energy
tensors on the brane.
For kinematics where all momenta are space-like off-shell with
$p \gg m$, the cubic interaction computed above dominates the
amplitude, which becomes strong at the scale \Eq{strongscale}.
Similarly, by studying the Feyman diagrams, one finds that \Eq{3int} leads
to a non-trivial 4-point scattering amplitude that 
violates unitarity at the scale $\Lambda$.

When we include quantum corrections, we expect all operators
consistent with symmetries to be generated, and we expect an
infinite number of terms that get strong at the scale $\La$.
Indeed, for the subset of logarithmically divergent graphs, we must include 
the associated operators
in order to be consistent with unitarity.%
\footnote{In other words,  logarithmic divergences 
correspond to the RG evolution, 
so that the coefficient of the corresponding operators cannot
be set to zero at all scales. 
Power divergent effects are not calculable and could consistently
be set to zero, for example by using dimensional regularization.}
These terms must be localized at the boundary, since the cutoff for
bulk interactions far from the boundary is $M_5 \gg \La_{\rm DGP}$.
These interactions must respect 4D Lorentz invariance and must
be local when written in terms of the geometrical quantities
$R_{\mu\nu\rho\si}(\ga)$ and $K_{\mu\nu}$. In order to zoom in on the strong interactions it is 
convenient to take a limit where $\Lambda= M_5^2/M_4$ is fixed and $M_4, M_5\to \infty$, so that also
$m=M_5^3/M_4^2 \to 0$. This  is the analogue of the $g\to 0$ with $f_\pi=m_V/g$ fixed limit
of massive non-abelian gauge theory. In this limit, the geometrical 
objects reduce to their linearized
approximation
\beq
m^{-2}R_{\mu\nu}&= \frac{\partial_\mu\partial_\nu \hat \pi}{\Lambda^3}+
O\left(\frac{m^2\partial \hat \pi\partial \hat\pi}{\Lambda^6}\right),\\
m^{-1}K_{\mu\nu}&= \frac{\partial_\mu\partial_\nu \hat \pi}{\Lambda^3}+
O\left(\frac{m^2\partial \hat \pi\partial \hat\pi}{\Lambda^6}\right)
\eeq
where $\hat \pi=\pi/M_4 m$ is the canonically normalized field.
By these equations we expect the terms that get strong at the
scale $\La$ to have the form 
\beq[counterterms]
\De \scr{L}_{\rm bdy} \sim \La^4 \left( \frac{\d}{\La} \right)^n
\left( \frac{R(\ga)}{m^2} \right)^p
\left( \frac{K}{m} \right)^q.
\eeq
Note that the theory becomes strongly coupled whenever the 4D curvature
is of order $m^2$. 
We will comment further on this point in Section 4 below.

Note also that the cubic interaction \Eq{3int} is nonlocal when written
in terms of geometrical quantities
(since $K \sim \d^2 \pi$ and $R \sim \d^2 \pi$).
This means that loops of $\pi$ fields should not renormalize this interaction,
and that the divergent part of loop diagrams involving this interaction
should be expressible as a  function of $\d^2 \pi$.%
\footnote{A full calculation should include loops of bulk fields as well.
However, the scaling argument shows the leading 
interactions at the scale $\La$ are expressible
in terms of the interactions of the $\pi$ field.}
This non-renormalization theorem follows simply by integration by
parts.
Consider any 1PI diagram with an external line coming from one
of the factors of $\pi$ with only one derivative.
Because the diagram is 1PI, both of the other $\pi$ factors attach
to internal lines.
We then have
\beq
\d^\mu \pi_{\rm ext} \d_\mu \pi_{\rm int} \Box_4 \pi_{\rm int}
= \d^\mu \pi_{\rm ext} \d_\nu \left[
\d_\mu \pi_{\rm int} \d_\nu \pi_{\rm int}
- \sfrac 12 \eta_{\mu\nu} \d^\rho \pi_{\rm int} \d_\rho \pi_{\rm int} \right],
\eeq
which is a function of $\d^2 \pi_{\rm ext}$ after integration by parts.
This gives a nice check of the consistency of this framework.

\section{Classical Instabilities}
In this section we study a classical solution to the DGP model
in which the stress-energy tensor on the brane satisfies the dominant
energy condition, yet the brane has negative energy from the 5D point of view.
When the boundary has the topology of $R^4$ it is difficult to define
the 5D energy, which is presumably infinite.
We therefore look for static solution where the spatial sections of the boundary
have topology $S^3$, the bulk is `outside' the $S^3$,
and the solution is $O(4)$ symmetric.
The geometry of the boundary therefore corresponds to a spatially compact
static cosmological solution, similar to the Einstein universe.

By Birkhoff's theorem, the metric outside the boundary is the 5D
Schwarzschild metric
\beq[SSmetric]
ds^2 = -f^2(r) dt^2 + \frac{dr^2}{f^2(r)} + r^2 d\Om_3^2,
\qquad
f = \sqrt{1 - \frac{R_{\rm S}^2}{r^2}}
\eeq
where $R_{\rm S}$ is the Schwarzschild radius.
The boundary is at a fixed value of $r > 0$ in these coordinates.
Because this solution asymptotes to 5D flat space at infinity, the energy
(mass)
of the solution is well-defined:
\beq[SSmass]
M = 32 \pi M_5^3 R_{\rm S}^2.
\eeq
For $R_{\rm S}^2 < 0$ such a solution has negative energy.
It is a negative-mass Schwarzschild solution with the naked singularity
cut out by the boundary.

The most general form of the stress-energy tensor on the brane
compatible with the symmetries is
\beq
T_{00} = -\rho \ga_{00},
\qquad
T_{ij} = + p \ga_{ij},
\eeq
where $\rho$ is the energy density and $p$ is the pressure.
We impose the equation of state
\beq
p = w \rho.
\eeq
We look for solutions satisfying the dominant energy condition,
which requires
\beq
\rho \ge 0,
\qquad
-1 \le w \le 1.
\eeq

The bulk Einstein equations are satisfied by the metric \Eq{SSmetric}.
The only additional equation that must be satisfied is
\beq[bdyequ]
4 M_4^2 \scr{G}_{\mu\nu}(\ga) - 4 M_5^3 (K_{\mu\nu} - \ga_{\mu\nu} K)
= T_{\mu\nu},
\eeq
where $\scr{G}_{\mu\nu}$ is the Einstein tensor
and $T_{\mu\nu}$ is a stress tensor on the boundary.
\Eq{bdyequ} follows simply from the variation of the full action 
with free boundary conditions.
(There is no junction equation in this approach since there is
no `other side' to the boundary.)
In the metric \Eq{SSmetric}, we can use \Eq{Kdef} to obtain
\beq
K_{\mu\nu} = \sfrac 12 f \d_r \ga_{\mu\nu}.
\eeq
The boundary equation \eq{bdyequ} then gives
\beq[Einsteintime]
4 M_4^2 \frac{3}{r^2} - 4 M_5^3 \frac{3 f}{r} &= \rho,
\\
\eql{Einsteinspace}
4 M_4^2 \frac{1}{r^2} - 4 M_5^3 \frac 1{r} \left( f + \frac 1{f} \right)
&= -w \rho.
\eeq
Note that when $M_5 = 0$ the 4D solution reduces to 
the standard Einstein static universe with $w=-1/3$, $\rho=12M_4^2/r^2$. 
When $M_4 = 0$, \Eq{bdyequ} is equivalent to the usual Israel junction
conditions, and \Eqs{Einsteintime} and \eq{Einsteinspace} have no
solutions satisfying the dominant energy condition.
(In fact, they have no solutions even if the stress energy tensor is allowed
to be written as a negative tension term plus a term satisfying the
dominant energy condition.)

The constraint $\rho \ge 0$ gives
\beq[weakineq]
r f \le \frac{1}{m}.
\eeq
Combining \Eqs{Einsteintime} and \eq{Einsteinspace} we obtain
\beq[massformula]
w = -\frac 13 + \frac{4 M_5^3}{\rho r f},
\eeq
which shows that $w \ge -1$ is satisfied whenever $\rho \ge 0$.
The condition $w \le 1$ gives
\beq[strongineq]
r f + \frac{r}{4 f} \le \frac 1m.
\eeq
Since this is clearly more restrictive than \Eq{weakineq}, this is
the only condition that needs to be checked.
For $M < 0$ ($R_{\rm S}^2 < 0$),
the \lhs of \Eq{strongineq} approaches $|R_{\rm S}|$ as $r \to 0$ and
increases monotonically with $r$, so we obtain
a solution for $|R_{\rm S}| < 1/m$.
This means that the energy cannot be made arbitrarily negative in this
model (see \Eq{SSmass}).

Note that the minimal 4D curvature of a negative energy solution is $\scr{O}(m^2)$.
For the critical zero energy solution $f \to 1$, and \Eq{strongineq}
implies $r \lsim 1/m$, so $\scr{G}_{\mu\nu} \gsim m^2$. However, 
\Eq{counterterms} shows that  such curvature is also the critical curvature 
where the derivative expansion of the effective
quantum field theory breaks down.
So the negative energy 
solutions to lie at the edge of the regime of validity of our theory.

Another way of arguing the same point is the following.
The instability appears for 4D energy density $\rho = \scr{O}(M_4^2 m^2$).
Significantly, this is also the energy density of a gravitational
source for which the cubic interaction term in \Eq{3int} becomes comparable 
to the kinetic term of the scalar $\pi$, defined in Section 3.
To see this, recall that  $\pi$ couples to the stress-energy 
tensor with strength $m$ (see \Eq{diag}).
Then, to linear order in the source, $\Box_4 \pi\sim T_\mu^\mu/M_4^2 m$. 
Substituting this estimate into the cubic interaction term \Eq{3int} we have 
\beq[m12]
\De \scr{L}_{\rm bdy} \sim T^\mu {}_\mu
\partial^\nu \pi \partial_\nu \pi.
\eeq
This term becomes of the same order as the $\pi$ kinetic term when
$T^\mu{}_\mu \sim M_4^2m^2$.
We conclude that the negative energy solutions appear only at the
edge of validity of the effective theory, and the theory with
cutoff of order $\La \sim (m^2 M_4)^{1/3}$ is safe from
instabilities.

\section{Curved Backgrounds}
A noteworthy aspect of massive gravity is that when propagating on a curved
background, it behaves very differently than in flat space. 
In AdS space, there is no vDVZ~\cite{vdvz} 
discontinuity~\cite{p}, while in dS a light massive graviton becomes a 
ghost~\cite{ds}. These unusual features find a simple explanation when massive
gravity is rendered covariant by adding a Goldstone vector~\cite{p2,AGS}
$A_\mu$. At linear order, $A_\mu$ appears in the combination
$h_{\mu\nu}-\bar{D}_{(\mu} A_{\nu)}$. Here $\bar{D}_\mu$ 
is the covariant derivative of 
the 4D background. The difference with flat space originates from the fact that
at nonzero cosmological constant $\Lambda$, the kinetic term of the 
strongly-interacting scalar mode $\pi$ inside $A_\mu$, $A_\mu=\bar{D}_\mu\pi$, 
receives a
contribution proportional to $-\Lambda$. This contribution suppresses the 
cubic interactions of $\pi$ when $\Lambda<0$, and makes the graviton a ghost
at small mass when $\Lambda>0$~\cite{AGS}.

In DGP, we find an analogous phenomenon.
We first give a general argument, then consider two important special cases:
one with a boundary with de Sitter geometry, the other a
Randall-Sundrum model with a DGP kinetic term on the boundaries.

\subsection{General Discussion}
We consider the linearized theory about a general 5D background metric
\beq
G_{MN} = \bar{G}_{MN} + H_{MN}.
\eeq
We generalize de Donder gauge to curved backgrounds
by adding the gauge fixing term
\beq
\De\scr{L}_{\rm bulk,gf} = -M_5^3 \bar{G}^{MN} F_M F_N,
\eeq
where
\beq
F_M = \bar{\nabla}^N H_{MN} - \sfrac 12 \bar{\nabla}_M H,
\eeq
where $\bar{\nabla}_M$ is the covariant derivative associated with
the background metric $\bar{G}_{MN}$.
The bulk equations of motion are then
\beq
\bar{\nabla}^2 \left( H_{MN} - \sfrac 12 \bar{G}_{MN} H \right) = 0.
\eeq

Following the discussion in the flat case, it is convenient to
parameterize the modes of $H_{MN}$ as follows.
For simplicity, we will give the discussion using Gaussian normal
coordinates for the background metric: $\bar{G}_{5\mu} = 0$,
$\bar{G}_{55} = 1$.
Instead of $H_{5\mu}| = h_{5\mu}$ and $H_{55}| = h_{55}$,
we use as boundary variables
\beq
\Xi_\mu | = \xi_\mu,
\qquad
\Xi_5 | = \pi,
\eeq
where $\Xi_M$ satisfies the bulk equation
\beq
\bar{\nabla}^2 \Xi_M = 0.
\eeq
This implies that the mode
$H_{MN} = \bar{\nabla}_{(M} \Xi_{N)}$
satisfies the bulk equations of motion as well as the de Donder
gauge fixing condition.
We then parameterize a general bulk fluctuation as
\beq[general]
H_{MN} = H_{MN}^{(2)} + \bar{\nabla}_{(M} \Xi_{N)} - \tilde{H}_{MN},
\eeq
where 
\beq
\bar{\nabla}^2 H^{(2)}_{MN} = 0,
\qquad
\bar{\nabla}^2 \tilde{H}^{(2)}_{MN} = 0,
\eeq
with boundary values
\beq
H_{\mu\nu}^{(2)} | = h_{\mu\nu},
\qquad
H_{5\mu}^{(2)} | = 0,
\qquad
H_{5 5}^{(2)} | = 0,
\eeq
and
\beq
\tilde{H}_{\mu\nu} | = \bar{\nabla}_{(\mu} \Xi_{\nu)} | 
= \bar{D}_{(\mu} \xi_{\nu)} + 2 \bar{K}_{\mu\mu} \xi_5,
\qquad
\tilde{H}_{5 \mu} | = 0,
\qquad
\tilde{H}_{55} | = 0,
\eeq
where $\bar{D}_\mu$ is the covariant derivative with respect to the
induced background metric $\bar{\ga}_{\mu\nu}$.
The $\tilde{H}_{MN}$ term in \Eq{general}
subtracts the contribution
of the `Goldstone' modes $\Xi_M$ to the fluctuations of the
induced boundary metric, which is then simply
$H_{\mu\nu}^{(2)}| = h_{\mu\nu}$, the `spin 2' mode.
This ensures that $\xi_\mu$ and $\pi$ as defined above do not appear in
the large DGP kinetic term.
We will see that in terms of these variables, the identification of the
strong degrees of freedom is more direct. 
(Indeed, the combination $N'_\mu$ that diagonalizes the kinetic term 
in \Eq{diag} in the flat case is precisely the `Goldstone' $\xi_\mu$.)

The dependence on $\xi_\mu$ can be obtained simply by noting that  under
\beq[symmetry]
h_{\mu\nu} \mapsto \bar{D}_{(\mu} \lambda_{\nu)},
\qquad
\xi_\mu \mapsto \xi_\mu + \lambda_\mu,
\qquad
\pi \mapsto \pi,
\eeq
the bulk fluctuation
\Eq{general} changes precisely by a residual gauge transformation $\Lambda_M$,
satisfying $\Lambda_\mu|=\lambda_\mu$, $\Lambda_5=0$.
\Eq{symmetry} is therefore a symmetry
of the quadratic boundary action.
Therefore, we can work out the action at $\xi_\mu = 0$ and restore the
dependence on $\xi_\mu$ by the substitution
\beq[xisub]
h_{\mu\nu} \to h_{\mu\nu} - \bar{D}_{(\mu} \xi_{\nu)}.
\eeq

We now compute the boundary action for the modes above.
For this we will need
\beq[d5curved]
\bar{\nabla}_5 H^{(2)}_{\mu\nu} | \simeq -\bar{\De} h_{\mu\nu},
\qquad
\bar{\nabla}_5 \Xi_\mu | \simeq - \bar{\De} \xi_\mu,
\qquad
\bar{\nabla}_5 \Pi | \simeq -\bar{\De} \pi,
\eeq
valid for modes with 4D wavelengths smaller than the scale of curvature,
where
\beq
\bar{\De} = \sqrt{-\bar{D}^2}.
\eeq
\Eq{d5curved} follows from the fact that for small-wavelength fluctuations
the curvature is irrelevant, so the result must reduce smoothly to the flat case.
We will see how this arises in
an explicit calculation in the next subsection.
The contribution to the boundary effective action
from the bulk Einstein action is then
\beq
\De\scr{L}_{\rm bdy, {\rm E}} &= -M_5^3 h^{\mu\nu}
\left[ \sqrt{-\ga} N ( K_{\mu\nu} - \ga_{\mu\nu} K )
\right]_{\rm linearized}
\\
\eql{bdycurveE}
&= -\sqrt{-\bar\ga} M_5^3 \bigl[
\sfrac 12 h^{\mu\nu} \bar\De h_{\mu\nu}
- \sfrac 12 h \bar\De h
+ h^{\mu\nu} \bar{D}_\mu \bar{D}_\nu \pi
- h \bar{D}^2 \pi + \cdots \Bigr],
\eeq
where we have omitted terms of order $\bar{K} h^2$ and
$\bar{K} h \bar{\De} \pi$ that are subleading at small 4D wavelengths.
The contribution from the bulk de Donder gauge fixing term is
\beq
\De\scr{L}_{\rm bdy, bulk\,gf} &= -M_5^3 \sqrt{-\bar\ga}
\left( H_{5M} F^M - \sfrac 12 H F^5 \right)
\\
&= M_5^3 \sqrt{-\bar\ga} \bigl[
-\sfrac 14 h \bar\De h - \pi (\bar{D}^\mu \bar{D}^\nu h_{\mu\nu}
- \bar{D}^2 h) 
\nonumber\\
\eql{bdycurvedD}
& \qquad\qquad\qquad\quad
+ 2 \pi ( \bar{K}_{\mu\nu} \bar{D}^\mu \bar{D}^\nu
- \bar{K} \bar{D}^2 ) \pi + \cdots \bigr]
\eeq
where
\beq
F_\mu &= \bar{D}^\nu (h_{\mu\nu} - 2 \bar{K}_{\mu\nu} \pi)
- \sfrac 12 \bar{D}_\mu (h - 2 \bar{K} \pi)
+ \scr{O}(\bar{K}^2 \pi, \bar{K} h),
\\
F_5 &= \sfrac 12 \bar{\De} (h - 2 \bar{K} \pi)
+ \scr{O}(\bar{K}^2 \pi, \bar{K} h),
\eeq
and we again omit terms that are subleading at small 4D wavelengths.
At this point, it is trivial to include the mode $\xi_\mu$ by the
substitution \Eq{xisub}.
Note that $\pi$ mixes with $h_{\mu\nu}$ only via the linearized
Ricci scalar, so there is no mixing between $\pi$ and $\xi_\mu$,
generalizing the result of flat case.

The full quadratic boundary action is therefore the sum of
\Eqs{bdycurveE} and \eq{bdycurvedD}.
In the small wavelength approximation, the only relevant change
comes from the last two terms in \Eq{bdycurvedD}.
To understand what they imply, we consider for simplicity a
maximally symmetric background, for which
\beq
\bar{K}_{\mu\nu} = C \bar{\ga}_{\mu\nu}.
\eeq
Adding the DGP kinetic term and diagonalizing the kinetic term for
$\pi$ and $h_{\mu\nu}$ we obtain the $\pi$ kinetic term
\beq[ghost]
\scr{L}_{\rm kin} = 3 M_4^2 m (m - 2C) \pi \bar{D}^2 \pi.
\eeq
Note that the $\pi$ mode becomes a ghost for $C = \sfrac 12 m$.
We will see below that $C > 0$ for de Sitter space (see \Eq{GKdSexplicit}).
Therefore, as in massive gravity, positive curvature increases the
strength of the strongest interactions, while negative curvature
decreases it.

One of the motivations for considering the DGP model is that it provides a
source of `dark energy' in the absence of 4D vacuum energy \cite{DDG}.
In this case, we can compute the relation between the constant $C$ introduced
above and the positive curvature of the present-day universe.
From \Eq{bdyequ}, we have
\beq[mmm6]
4M^2_4 \scr{G}_{\mu\nu}(\ga)
= T_{\mu\nu} - 12M^3_5 C \gamma_{\mu\nu} \simeq
T_{\mu\nu} - 4M^2_4 \la_{\rm now} \ga_{\mu\nu}.
\eeq
This shows that $C$ is positive, hence $\pi$ is a ghost, for sufficiently 
small $m$.
In the next subsection, we will see that in a DGP model with no cosmological
constant in the bulk, $\pi$ is a ghost in the regime where 4D vacuum energy does
not contribute to the 4D curvature.
More generally, we expect that positive 4D curvature decreases the
strength of the $\pi$ kinetic term, so that it makes the interactions of $\pi$
even stronger than on a Minkowsky 4D background. 
Conversely, a negative 4D curvature weakens the $\pi$ self-interactions,
as in massive gravity.
When $m$ is much smaller than the curvature $|C|$, it even eliminates the vDVZ discontinuity already at 
linear order~\cite{p}.

\subsection{Explicit Calculation: de Sitter Space}
We now consider the important special case of 4D de Sitter space in a
DGP model with vanishing bulk cosmological constant.
The solution has very simple 5D geometry:
the bulk is flat, and the boundary is at
\beq
\eta_{\mu\nu} x^\mu x^\nu + y^2 = L^2
\eeq
in Cartesian coordinates.
It is more convenient to use coordinates where the background metric is
\beq[dSbackmet]
ds^2 = dr^2 - r^2 d\tau^2 + r^2 \cosh^2\tau\, d\Om_3^2.
\eeq
The boundary is at $r = L$ in these coordinates.
The boundary equation is
\beq
4 M_4^2 \scr{G}_{\mu\nu}(\bar\ga)
- 4 M_5^2 \left[ \bar{K}_{\mu\nu} - \bar{\ga}_{\mu\nu} \bar{K} \right]
= - V_0 \bar{\ga}_{\mu\nu},
\eeq
where $V_0 > 0$ is the vacuum energy on the boundary.
It is straightforward to work out
\beq[GKdSexplicit]
\scr{G}_{\mu\nu}(\bar\ga) = -\frac{3}{r^2} \bar{\ga}_{\mu\nu},
\qquad
\bar{K}_{\mu\nu} = +\frac{1}{r} \bar{\ga}_{\mu\nu}.
\eeq
This gives the relation between the 4D vacuum energy and the
curvature:
\beq
\frac{V_0}{12 M_4^2} L^2 + m L = 1.
\eeq
For $V_0 / M_4^2 \gg m^2$, this gives the usual relation between
vacuum energy and curvature, but for $V_0 / M_4^2 \ll m^2$ we get
$L \simeq 1/m$ \cite{DDG}.
It is interesting that in this regime the 4D curvature is independent
of the vacuum energy, but from \Eq{ghost} we see that the Goldstone is always
a ghost in this regime.

We now consider the explicit calculation of the boundary action.
To understand the strong interactions, it is sufficient to consider
the scalar modes
\beq
H_{MN} = \bar{G}_{MN} \Phi
+ ( \bar{n}_M \bar{\nabla}_N \Si + \bar{n}_N \bar{\nabla}_M \Si )
+ \bar{n}_M \bar{n}_N \Om,
\eeq
where $\bar{n}_M$ is the normal vector
$\bar{n}^5 = \bar{n}_5 = 1$,
$\bar{n}^\mu, \bar{n}_\mu = 0$.
A useful relation is
\beq
\bar{\nabla}_M \bar{n}_N = \frac 1r \bar{\ga}_{MN},
\eeq
where
\beq
\bar{\ga}_{MN} = \bar{G}_{MN} - \bar{n}_M \bar{n}_N
\eeq
is the induced metric on surfaces of constant $r$.%
\footnote{We can think of $r$ as a bulk scalar.
Geometrically, it is the proper distance of any point in the
bulk to the boundary.}
The de Donder equations of motion for the scalar modes defined
above are
\beq
\bar{\ga}_{\mu\nu} \left( \bar{\nabla}^2 \Phi + \frac 2{r^2} \Om \right)
+ \frac 4r \bar{\nabla}_\mu \bar{\nabla}_\nu \Si &= 0,
\\
\bar{\nabla}_\mu \left[
\bar{\nabla}^2 \Si
+ \frac 2r \bar{\nabla}_r \Si
- \frac 4{r^2} \Si
+ \frac 2r \Om \right] &= 0,
\\
\bar{\nabla}^2 \Om + \bar{\nabla}^2 \Phi
+ 2 \bar{\nabla}_r \left( \bar{\nabla}^2 \Si - \frac 4{r^2} \Si \right)
- \frac 8{r^2} \Om &= 0.
\eeq
In our coordinates
\beq
\bar{\nabla}^2 \Phi = \left[ \d_r^2 + \frac 4r \d_r 
+ \frac 1{r^2} \hat{\Box}_4 \right] \Phi,
\eeq
where $\hat{\Box}_4$ is the Laplacian on $S^3$.
Since $\hat{\Box}_4$ is independent of $r$, we can treat it as a
parameter when solving the equations.
(More formally, we could expand in eigenstates of $\hat{\Box}_4$.)

The behavior of the solutions is very easy to understand once
we notice that they are homogeneous in $r$.
The solutions therefore have the form
\beq
\Phi = \left( \frac rL \right)^A \phi,
\qquad
\Si = \left( \frac rL \right)^{A + 1} \si,
\qquad
\Om \left( \frac rL \right)^A \om,
\eeq
where $A$ depends on $\hat{\Box}_4$.
The solution is very simple for fluctuations with
$\Box_4 \gg 1/L^2$ on the boundary.
(Recall that $L$ is the size of the 4D universe!)
For these fluctuations $\hat{\Box}_4 \gg 1$ and the leading
terms in the equations that determine $A$ are simply
\beq
A^2 + \hat{\Box}_4 = 0,
\eeq
with solution
$A = \pm \sqrt{-\hat{\Box}_4}$.
Good behavior at infinity (away from the boundary)
requires the negative solution.
Since $\hat{\Box}_4$ is the only expansion parameter, the
corrections are
\beq
A = -\sqrt{-\hat{\Box}_4} \left[ 1 + \scr{O}(1/ \hat{\Box}_4) \right],
\eeq
and so on the boundary
\beq
A | = - L \bar{\De} + \scr{O}(1/L \bar{\De}).
\eeq
Note that the corrections to $A$ are smaller than the $1/L$ curvature
corrections described in the previous subsection.

The conclusions depend only on the fact that the equations are
second order and homogeneous in $r$, and so hold for general
polarization states.
This shows that
\beq
\d_r H_{MN} | = -\bar{\De} H_{MN} + \scr{O}(H / L^3 \bar{\De}).
\eeq
The remainder of the calculation follows the previous subsection
line by line.

\subsection{DGP and Randall-Sundrum}
It is instructive to apply the results of the previous section 
to the Randall-Sundrum (RS) model~\cite{rs} with a DGP boundary term added.
At the boundaries, we have
\beq
\bar{K}_{\mu\nu} = \pm L \bar{\ga}_{\mu\nu},
\eeq
where $L = 1/k$ is the bulk AdS curvature length, and
the $+$ ($-$) sign corresponds to the IR (Planck) brane.

We first consider adding a DGP kinetic term on the Planck brane.
As long as $k \gsim m$, \Eq{ghost} shows that the scale of strong
interactions is $(M_5 k)^{1/2} \gsim k$.
We conclude that the DGP kinetic term
does not lead to new strong interactions within the original
regime of validity of the RS model.
This is consistent with the holographic interpretation of the model
as a 4D conformal field theory (CFT) coupled to gravity.
The DGP kinetic term simply corresponds to a large coefficient for
the gravity kinetic term in the UV, large enough to dominate the 
induced contribution to the Planck scale from the CFT.
Note that the extrinsic curvature term in the boundary effective
action is crucial for obtaining this result.

We now consider adding a DGP kinetic term to the IR (or `TeV') brane.
In this case, \Eq{ghost} tells us that the theory has a ghost for
$k > \sfrac 12 m$.
In fact, this instability comes from the radion itself becoming a
ghost.
To see this in a simple way, we consider the limit where the Planck
brane is pushed to the boundary of AdS and the IR brane is at fixed
position $y = 0$.
In this limit the 4D zero mode graviton is decoupled from the physics
on the IR brane.
The only zero mode is the radion $\phi$, which can be parameterized by
\cite{Charm}
\beq
ds^2 = e^{2 k y} \left[ 1 + 2 \phi(x) e^{-2 k y} \right]
dx_\mu dx^\mu + \left[ 1 - 2 \phi(x) e^{-2 k y} \right]^2 dy^2.
\eeq
Its kinetic term is
\beq
\scr{L}_{\rm kin} =
6 \left( \frac{M_5^3}{k} - 2 M_4^2 \right) \phi \Box_4 \phi,
\eeq
in agreement with \Eq{ghost}.

This is interesting because it prevents a geometrical construction of
a model with an isolated massive spin 2 particle.
This model has only massive spin 2 KK modes, and when
$M_4^2 \gg M_5^3 / k$ one finds that the lightest spin 2 mode
has an anomalously small mass of order $(k m)^{1/2}$,
while the remaining massive KK modes have mass of order $k$.
If it were not for the instability, discussed above, for
$m \ll k$ this model would reduce to an effective theory of a
single massive graviton (and a radion) below the scale $k$.

\newpage
\section{Conclusions}
We now summarize our results.
First, we showed that the DGP model has strong interactions at distances
shorter than $\la_3 \sim (\la_{\rm DGP}^2 / M_4)^{1/3}$.
The strong interactions are due to a scalar `Goldstone' mode that
obtains a kinetic term only by mixing with the transverse graviton
polarizations, similar to massive gravity.
The longitudinal Goldstone has a geometrical interpretation as a 
brane bending mode that keeps the induced metric on the brane fixed.
Second, we showed that there are classical instabilities in the DGP
model in the form of negative energy solutions.
These solutions are at the edge of validity of the effective field
theory with UV cutoff at the scale $\la_3$, giving further support
to the conclusion that new physics is required at this scale.
Finally, we considered the strong interactions in the presence of
4D and/or 5D curvature.
We showed that positive (de Sitter) sign curvature makes the model
more strongly interacting, and makes the strong mode a ghost for
sufficiently large curvature.
We also investigated the effect of a DGP kinetic term in the Randall-Sundrum
model.

We conclude with some comments on the solution of the vDVZ~\cite{vdvz} 
discontinuity problem suggested in \Ref{ddgv}.
There, it was shown that around a classical source of Schwarzschild radius 
$R_{\rm S}$, a careful resummation of non-linear effects restores the 
phenomenologically correct Schwarzschild solution
below a distance $R_* \sim (R_{\rm S} / m^2)^{1/3}$.
Now, the curvature of the Schwarzschild metric is of order
$R_{\rm S}/r^3$, so that at the distance $R_*$ the curvature is of order
$m^2$. 
From inspection of \Eq{counterterms} we find that this is the 
critical curvature at which the quantum effective field theory breaks down.
Stated otherwise, at the scale $R_*$, the effective quantum expansion parameter
$\ep = \partial^2\hat \pi/\Lambda^3$
becomes order 1.
Therefore, any statement about the behavior of the
field at distances smaller than $R_*$ requires 
knowledge (or assumptions) about the UV completion of the $\pi$ sector.

One possibility is to assume that when $\ep \gg 1$ the infinite series of 
counterterms saturates and the quantum correction to the effective action
stays of the order of the result at $\ep \sim 1$.
In this case, the contribution from the counterterms is suppressed compared
to the tree-level contribution (which has fewer derivatives)
by $\sim 1/(M_4 R_{\rm S})^2$.
In this scenario, there is a range of scales where the effect of \Ref{ddgv}
works.
In this case, one still has to find a way to avoid the instabilities associated
with curvature of order $m^2$, found above.
Moreover, this does not rescue the model phenomenologically, since at least
at the length scale $\la_3 \sim 1000$~km, gravity becomes sensitive to the
details of the UV completion.

\newpage
\section*{Acknowledgements}
We would like to thank Raman Sundrum for collaboration in the early stages of
this project.
We also thank Nima Arkani-Hamed, Thibault Damour, Gia Dvali, Gregory Gabadadze, Rob Leigh and
Valery Rubakov for useful and stimulating discussions.
M.A.L. was supported by NSF grant PHY-0099544.
M.P. conducted a significant part of research related to this paper while
on sabbatical at the Institute for Advanced Studies, Princeton NJ, 
and the Scuola Normale Superiore, Pisa, Italy. He would like to thank their 
gracious hospitality and support.
M.P. is supported in part by NSF grant  PHY-0070787.
R.R. would like to thank the 
Scuola Normale Superiore for hospitality during this project.


\end{document}